\documentclass[creativecommons,sharealike]{eptcs}
\providecommand{\event}{Working Formal Methods Symposium 2024} 

\usepackage{iftex}

\usepackage{amsmath}

\usepackage{csquotes}
\usepackage{hyperref}
\usepackage[capitalise]{cleveref}

\usepackage{graphicx}
\usepackage{color}
\usepackage{tikz-cd}

\usepackage{titlesec}
\titlespacing*{\paragraph} {1em}{0.3em}{0.3em}

\usepackage{subcaption}

\usepackage{bussproofs}
\usepackage{simplebnf}

\usepackage[inline]{enumitem}
\setlist[enumerate,1]{label={(\arabic*)}}

\usepackage{multicol}

\ifpdf
  \usepackage{underscore}         
  \usepackage[T1]{fontenc}        
\else
  \usepackage{breakurl}           
\fi

\newcommand*{\defEq}{\mathrel{:=}}

\newcommand*{\typed}[2]{#1 \mathrel{:} #2}

\newcommand*{\typeGroup}[2]{#1\left[#2\right]}
\newcommand*{\groupcombine}{\cup}

\newcommand*{\actionsilent}{\tau}
\newcommand*{\actionin}[2]{#1(#2)}
\newcommand*{\actionout}[2]{\overline{#1}\mathord{\langle}#2\mathord{\rangle}}
\newcommand*{\actionboundout}[3]{(\mathrm{\nu} \, \typed{#2}{#3})\overline{#1}\left\langle#2\right\rangle}

\newcommand*{\processnull}{\mathbf{0}}
\newcommand*{\processpar}{\mid}
\newcommand*{\processchoice}{\mathbin{+}}
\newcommand*{\processrep}{\mathord{!}}
\newcommand*{\processname}[2]{(\mathrm{\nu} \, \typed{#1}{#2})}
\newcommand*{\processsilent}{\tau.}
\newcommand*{\processin}[3]{\actionin{#1}{\typed{#2}{#3}}.}
\newcommand*{\processout}[2]{\actionout{#1}{#2}.}
\newcommand*{\processfullconditional}[4]{\left[#1 \mathbin{=} #2\right]\left(#3\mathbin{;}#4\right)}
\newcommand*{\processconditional}[4]{\left[#1 \mathrel{#2} #3\right] #4}
\newcommand*{\processeqconditional}[3]{\processconditional{#1}{=}{#2}{#3}}
\newcommand*{\processuneqconditional}[3]{\processconditional{#1}{\neq}{#2}{#3}}

\newcommand*{\systemnull}{\mathbf{0}}
\newcommand*{\systempar}{\mathbin{\parallel}}
\newcommand*{\systemgroup}[2]{#1\left[#2\right]}
\newcommand*{\systempurpose}[3]{#1 : #2\left[#3\right]}
\newcommand*{\systemname}[2]{(\mathrm{\nu} \, \typed{#1}{#2})}

\DeclareMathOperator*{\freenames}{fn}
\DeclareMathOperator*{\boundnames}{bn}

\newcommand*{\subst}[2]{\left[#2 \mathbin{\mathtt{:=}}#1\right]} 

\newcommand*{\shiftdown}[2]{\left[\mathtt{shiftdown}\,#1\, #2\right]}

\newcommand*{\transition}[1]{\mathrel{\xrightarrow{#1}}}
\newcommand*{\transitionsilent}{\transition{\actionsilent}}
\newcommand*{\transitionin}[2]{\transition{\actionin{#1}{#2}}}
\newcommand*{\transitionout}[2]{\transition{\actionout{#1}{#2}}}
\newcommand*{\transitionboundout}[3]{\transition{\actionboundout{#1}{#2}{#3}}}

\def\DEnv/{$\Delta$-Environment}
\def\DEnvs/{$\Delta$-Environments}

\def\ThInt/{$\Theta$-Interface}
\def\ThInts/{$\Theta$-Interfaces}

\def\GEnv/{$\Gamma$-Environment}
\def\GEnvs/{$\Gamma$-Environments}

\def\piCalculus/{$\pi$-calculus}
\def\alphaEquivalence/{$\alpha$-equivalence}
\def\alphaEquivalent/{$\alpha$-equivalent}

\newcommand*{\convertToPrivacyCalculus}[1]{\left\|#1\right\|}
\newcommand*{\tokenType}{\mathtt{Token}}
\newcommand*{\tokenValue}{\mathtt{t}}

\title{Converting BPMN Diagrams to Privacy Calculus}
\author{
Georgios V. Pitsiladis\qquad\qquad Petros S. Stefaneas
\institute{Department of Mathematics\\ School of Applied Mathematical and Physical Sciences\\National Technical University of Athens\\9 Iroon Polytechniou Str., 15772 Zografou, Greece}
\email{\quad gpitsiladis@mail.ntua.gr \quad\qquad petros@math.ntua.gr}
}
\def\titlerunning{Converting BPMN Diagrams to Privacy Calculus}
\def\authorrunning{G. V. Pitsiladis, P. S. Stefaneas}
\begin{document}
\maketitle

\begin{abstract}
	The ecosystem of Privacy Calculus is a formal framework for privacy comprising (a) the Privacy Calculus, a Turing-complete language of message-exchanging processes based on the  \piCalculus/, (b) a privacy policy language, and (c) a type checker that checks adherence of Privacy Calculus terms to privacy policies.
	BPMN is a standard for the graphical description of business processes which aims to be understandable by all business users, from those with no technical background to those implementing software.
	This paper presents how (a subset of) BPMN diagrams can be converted to Privacy Calculus terms, in the hope that it will serve as a small piece of larger workflows for building privacy-preserving software. 
	The conversion is described mathematically in the paper, but has also been implemented as a software tool.
\end{abstract}

\section{Introduction}\label{sect:introduction}

The main motivation of this paper is that it might serve as a first version of a piece of a larger workflow for building privacy-preserving software.

In order to trust that some piece of software is privacy-preserving, this must somehow be proved formally; in other words, privacy protection needs to be considered as a formal specification (formalised privacy policies) complemented by tools able to decide adherence of programs to policies.

The Privacy Calculus ecosystem has been introduced in~\cite{kouzapas2014} to tackle these considerations; it was further developed in~\cite{kouzapas2015,kokkinofta2014,pitsiladis2016,pitsiladis2016a,kouzapas2017,vanezi2020a}. Privacy Calculus is a variation of the \piCalculus/, a Turing-complete language describing parallel processes sharing messages.
It is accompanied by a privacy policy language, which gives the ability to grant \emph{permissions} (read, write, disclose, store, etc.) to \emph{users} or \emph{groups} (forming a hierarchy) for specific \emph{purposes}\footnote{The notion of purpose is inherent in privacy protection. This has been argued in the literature regarding privacy, but has also been acknowledged in practice by legislation: purpose of data processing is a fundamental notion in GDPR.}.
The ecosystem is completed by a type checker for checking compliance of Privacy Calculus terms to privacy policies written in the aforementioned formal language. 

Although some tools for working with the Privacy Calculus ecosystem have been created~\cite{pitsiladis2018,vanezi2020}, the ecosystem is still quite abstract, far from everyday practice. One way to bridge this gap is to create conversions between higher-level frameworks and Privacy Calculus. This is where BPMN might fruitfully enter the discussion.

The aim of Business Process Model and Notation (BPMN) is to serve as a standard for the graphical depiction of business processes, enhancing intra- and inter-organisational communication and interoperability. It is high-level enough to be understandable by audiences with minimal technical background, however it can be quite detailed and (in its full generality) even automatically executable.

This paper is an exploration of how the most basic elements of BPMN can be converted to Privacy Calculus terms with the hope that eventually, a workflow such as the following could be feasible:
\begin{enumerate*}
	\item describe a business process in BPMN,
	\item convert it to Privacy Calculus,
	\item specify a privacy policy, ideally by converting it from some high-level framework to the formal privacy policy language,
	\item obtain (e.g. with the Maude tool presented in~\cite{pitsiladis2018}) a proof that the business process adheres to the policy.
\end{enumerate*}

The rest of this paper is organised as follows: \cref{sect:bpmn} reviews basic notions of BPMN, \cref{sect:privacy-calculus} contains some basic definitions of the Privacy Calculus, and \cref{sect:conversion} discusses how BPMN diagrams (or rather, a subset of them) can be converted to Privacy Calculus terms and presents a tool that automates the said conversion; \cref{sect:conclusion} contains some concluding remarks.

\section{Business Process Model and Notation}\label{sect:bpmn}

BPMN defines, both syntactically and semantically, a multitude of graphical elements. 
These elements can be combined into diagrams.
Three kinds of diagrams are possible: Collaborations, Processes, and Choreographies (\cite[Section~1.1]{bpmn}); here, only the first two will be considered.

\begin{figure}[!htb]
\begin{subfigure}{\textwidth}
\begin{center}
\includegraphics[width=0.4\textwidth]{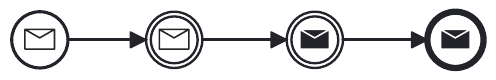}
\end{center}
\caption{A Process with only Message Events as Flow Nodes. From start to end, it contains a Message Start Event, a Message Intermediate Catch Event, a Message Intermediate Throw Event, and a Message End Event.}
\label{fig:diagram-messages}
\end{subfigure}
\begin{subfigure}{\textwidth}
\begin{center}
\includegraphics[width=0.55\textwidth]{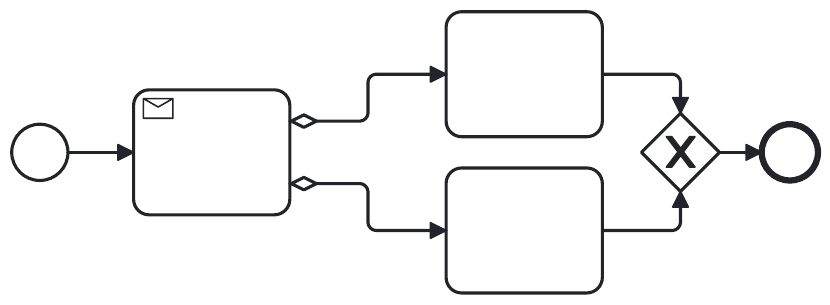}
\end{center}
\caption{A Process with some Conditional Flows (recognised by the diamond symbol at their start). The Start Event is followed by a Receive Task. Depending on the conditions, one or both of the following two Abstract Tasks are triggered; here, it is assumed that the conditions are such that only one can be fulfilled. An Exclusive Gateway combines the two alternative paths and channels the flow to the End Event.}
\label{fig:diagram-conditional}
\end{subfigure}
\caption{Two diagrams of BPMN Processes.}
\end{figure}

Processes can be public or private. Private Processes model activities within an organisation: they can be defined at a so high level of detail as to be executable; otherwise, they serve for documentation purposes. Public Processes are non-executable and show activities of multiple Participants, documenting their interaction and hiding (parts of) actions internal to Participants~\cite[Section~7.2.1]{bpmn}. Here, since the interest lies on data protection among multiple stakeholders (the data subject and at least one data processing entity), mostly public Processes will be considered. \cref{fig:diagram-conditional,fig:diagram-messages} are examples of Processes.

A Collaboration contains two or more Participants and its purpose is to depict the interactions among them~\cite[Section~7.2.1]{bpmn}. Each Participant is depicted as a Pool which may be empty or contain a Process diagram~\cite[Table~7.1]{bpmn} (at most one process can be private, in which case it may be drawn outside of a Pool~\cite[Section~9.3]{bpmn}). Pools can also be divided in Lanes and/or have multiple instances, but these features will not be considered here. \cref{fig:diagram-looping-subprocess} is an example of a Collaboration with two Pools.

There are five categories of graphical elements~\cite[Section~7.3]{bpmn}: flow objects (Events, Activities, Gateways), data (Data Objects, Data Stores), connecting objects (Sequence Flows, Message Flows, Associations, Data Associations), swimlanes (Pools, Lanes), and artifacts (Text Annotations, element Groups). Here, only flow objects, Flows, and Pools will be considered; the main characteristics of Events, Activities, Gateways, and Flows will be presented in \cref{sect:bpmn-events,sect:bpmn-activities,sect:bpmn-gateways,sect:bpmn-flows}. Data Objects and Data Stores will not be considered, since the version of Privacy Calculus employed here cannot deal with them properly. Messages will only be considered indirectly, because they are not supported by the \href{https://bpmn.io}{bpmn.io} diagram editor; when needed, they will be considered as available externally to the BPMN modelling.

In order to understand how control flows within a diagram, the concept of \emph{tokens} is employed in lieu of semantics; in the words of \cite[Section~7.2]{bpmn}, \enquote{A token is a \emph{theoretical} concept that is used as an aid to define the behavior of a Process that is being performed. The behavior of Process elements can be defined by describing how they interact with a token as it \enquote{traverses} the structure of the Process.}. In short (and with many details omitted), Start Events generate tokens, End Events consume them, and the other elements redirect, multiply, or merge them appropriately. Tokens will be instrumental for the conversion to Privacy Calculus.

\subsection{Events}\label{sect:bpmn-events}

There are three types of Events\footnote{BPMN also defines Events at the boundaries of Activities~\cite[Section~10.5.4]{bpmn}, but these will not be considered here.} based on \emph{when} they affect the flow of a process: Start, Intermediate, and End. There are also multiple types of events depending on \emph{how} the affect the flow: here, only Message Events (and Start/End events with no information as to how they affect the flow) will be considered. Every Event either catches or throws (but not both): Start Events always catch, End Events always throw, Intermediate Message Events may do either. \cite[Section~7.3.2]{bpmn}

Contrary to~\cite[Section~10.5.2 and~Section~7.2.1]{bpmn}, which allow leaving Start/End events implicit for simplicity, this paper requires that Processes (and Sub-Processes) must always start with one or more Start Events and that each path of a Process (or Sub-Process) must terminate at an End Event.
This affects expressiveness only minimally; moreover, in future treatments, \enquote{phantom} Start/End Events, connected to the \enquote{initial} and \enquote{final} Flow Nodes, could be added if none are provided.

Naturally, Start Events have no incoming Sequence Flows and End Events have no outgoing Sequence Flows. In order to accommodate implicit Start/End Events, BPMN~(\cite[Section~10.5.2]{bpmn}) allows this for other Flow Nodes as well. Here, the only Flow Nodes that will be permitted not to have an incoming Sequence Flow are Start Events; dually, the only Flow Nodes that will be permitted not to have an outgoing Sequence Flow are End Events and Sub-processes. 

\subsection{Activities}\label{sect:bpmn-activities}

Activities are divided into Tasks and Sub-Processes. Tasks are atomic (as far as the modelling is concerned), while Sub-Processes are compound~\cite[Section~7.3.2]{bpmn}. 

\begin{figure}[!htb]
\begin{center}
\includegraphics[width=0.85\textwidth]{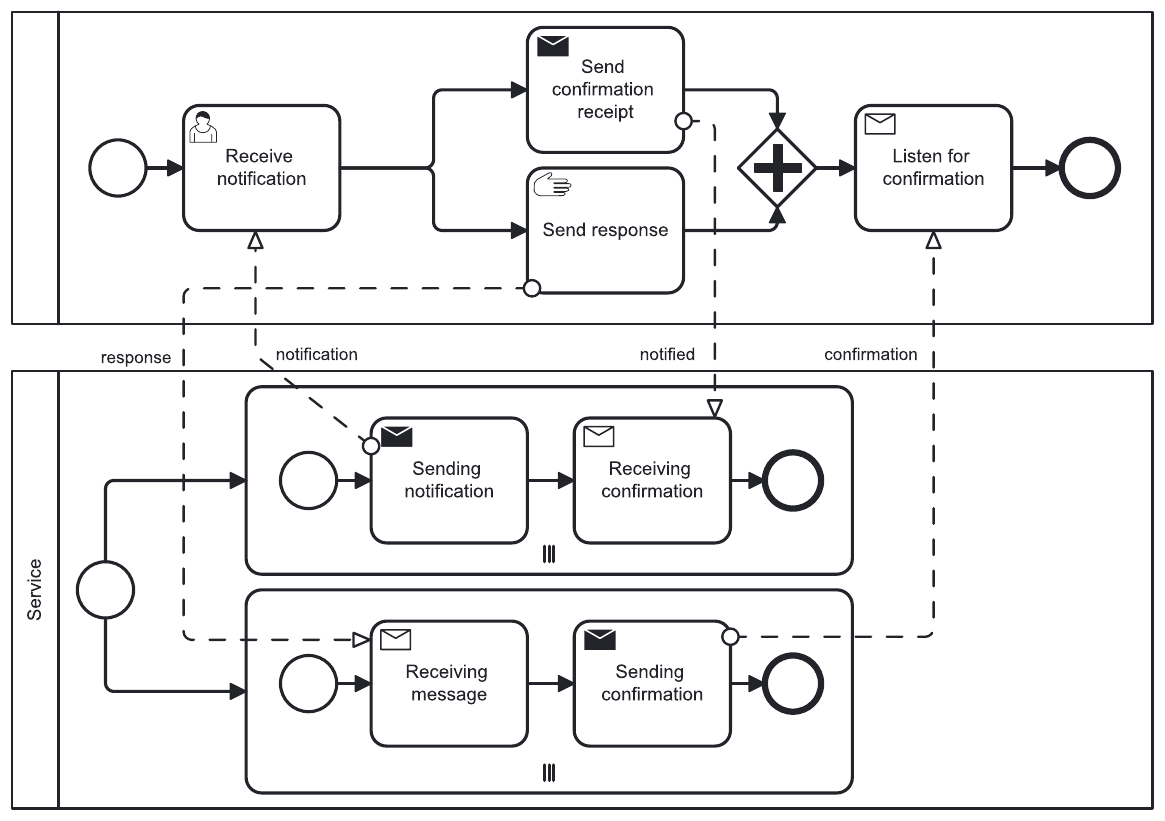}
\end{center}
\caption{A BPMN diagram depicting a Collaboration. Here, the \enquote{Service} Pool has two Sub-Processes. These Sub-Processes have multiple (parallel) instances, indicated by the parallel lines at their bottom.}
\label{fig:diagram-looping-subprocess}
\end{figure}

\paragraph{Tasks:} 
A Task is an Activity which represents some action not broken down to more detail, hence considered atomic (in fact, it might be cancelled in mid-execution through the Compensation or other mechanisms of BPMN, but this is not considered here). There are many types of Tasks, including a generic one (Abstract Tasks). 
Apart from Abstract Tasks, this paper is mainly interested in Send Tasks (e.g. \enquote{Send confirmation receipt} in \cref{fig:diagram-looping-subprocess}), which send Messages to other Participants, and Receive Tasks (e.g. \enquote{Listen for confirmation} in \cref{fig:diagram-looping-subprocess}), which receive Messages from other Participants.
Among the rest types of Tasks are User Tasks (e.g. \enquote{Receive notification} in \cref{fig:diagram-looping-subprocess}), which are executed by humans with the aid of automated systems, and Manual Tasks, which are executed by humans manually (e.g. \enquote{Send response} in \cref{fig:diagram-looping-subprocess}; imagine that the response is sent via traditional mail).

Some simplifying conventions (limiting the expressiveness of our tool) will be made. Contrary to~\cite[Section~10.3]{bpmn}, here it will be assumed that every Task has at most (hence, exactly) one incoming Sequence Flow, at most one incoming Message Flow, and at most one outgoing Message Flow.
Also, looping and multiplicity of Tasks will not be considered here.

\paragraph{Sub-Processes:}

BPMN defines some special types of Sub-Processes; here, however, we will only be interested in those that are just Processes within Processes (Embedded Sub-Processes). Examples can be seen in \cref{fig:diagram-looping-subprocess}.
Sub-Processes may have parallel multiplicity, i.e. multiple copies of a Sub-Process may run in parallel (looping or sequential multiplicity are also options in BPMN, but will not be considered here).
Recall that here we require Sub-Processes to always contain at least one Start and one End Event.
As for Tasks, contrary to~\cite[Section~10.3]{bpmn}, here it will be assumed that every Sub-Process has at most (hence, exactly) one incoming Sequence Flow. 

\subsection{Flows}\label{sect:bpmn-flows}

Flows are drawn as arrows. There are two kinds of flows: Sequence Flows (solid arrows) and Message Flows (dashed arrows).

\paragraph{Sequence Flows:}
Sequence Flows model the flow of control within a Process.

An Uncontrolled Flow (i.e. a Normal Flow not connected to some Gateway and not affected by Conditions) is the most basic kind of Flow, representing the order of execution of the elements it connects.

Conditional Flows (examples can be seen in \cref{fig:diagram-conditional}) are only activated if their corresponding Condition is met. Here, Conditional Flows will be considered only when they are outgoing from a Receive Task (their Condition shall then pertain to the received message).
Default Flows will not be considered.
 
Non-Normal Sequence Flows (pertaining to Exceptions and Compensations), as well as looping via \enquote{backwards} sequence flows, will not be considered here.

Of course, the restrictions on Flows heavily impact the expressiveness of the diagrams our tool can handle. However, the supported elements are already expressive enough to be considered in this version.
 
\paragraph{Message Flows:}
Message Flows depict the flow of messages between Participants in a Collaboration~\cite[Section~7.3.2]{bpmn} (see \cref{fig:diagram-looping-subprocess}).
According to~\cite[Chapter~10]{bpmn}, \enquote{All Message Flows must connect two separate Pools. They may connect to the Pool boundary or to Flow Objects within the Pool boundary.}. 
In this paper, each Message Flow must have a Flow Object as either source or target.
Contrary to~\cite[Section~7.6.2]{bpmn}, here it will be assumed that Sub-Processes per se have neither incoming nor outgoing Message Flows. Flow Nodes \emph{within} a Sub-Process will support Message Flows as usual.

\subsection{Gateways}\label{sect:bpmn-gateways}

Gateways control the convergence and divergence of Sequence Flows. Here, only Parallel and Exclusive Gateways will be considered (the latter only in their converging form).
A Gateway must have multiple incoming Sequence Flows or multiple outgoing Sequence Flows (or multiple of both, which is not recommended)~\cite[Section~10.6]{bpmn}; recall that here it is also required to have at least one of each.
\blockcquote[Section~10.6]{bpmn}{Gateways do not represent \enquote{work} being done and they are considered to have zero effect on the operational measures of the Process being executed (cost, time, etc.).}

Parallel Gateways (e.g. the one synchronising the Tasks \enquote{Send confirmation receipt} and \enquote{Send response} in \cref{fig:diagram-looping-subprocess}) create and synchronise parallel paths: that is, when multiple Flows are outgoing from a Parallel Gateway, all of them will be executed in parallel; dually, when multiple Flows are incoming to a Parallel Gateway, it will wait until all of them are executed before activating its output Flows.

According to~\cite[Section~10.6.2]{bpmn}, \enquote{A converging Exclusive Gateway \textins{i.e. one with multiple incoming Sequence Flows and one outgoing Sequence Flow} is used to merge alternative paths. Each incoming Sequence Flow token is routed to the outgoing Sequence Flow without synchronization.}. Here, we will only consider cases where the incoming Sequence Flows to the Exclusive Gateway are alternative, i.e. that at most one of them can be triggered. An example of an Exclusive Gateway can be seen in \cref{fig:diagram-conditional}.

\section{Privacy Calculus}\label{sect:privacy-calculus}

The Privacy Calculus is a typed variant of the \piCalculus/ introduced in~\cite{kouzapas2014} and further developed in~\cite{kouzapas2015,kokkinofta2014,pitsiladis2016,pitsiladis2016a,kouzapas2017,vanezi2020a}. The version of the Privacy Calculus presented here is the one of~\cite{pitsiladis2018}, with the addition of the Choice and Silent operators which are standard in \piCalculus/ and can be introduced in the tool of~\cite{pitsiladis2018} with minimal effort (amounting to the addition of two transition rules, i.e. \eqref{pi:Choice} and \eqref{pi:Silent} of \cref{fig:pi-transitions}, and two simple typing rules).

Assume the following basic sets of entities: 
\begin{enumerate*}
\item an infinitely countable set of channel names (ranged over by $x$, $y$, $z$, $a$, $b$), 
\item a set of basic types, 
\item a set of purposes (ranged over by $u$),
\item a set of groups, (ranged over by $G$), split into a set of users, (ranged over by $U$) and a set of roles (ranged over by $R$); for any two groups $G_1, G_2$, their union $G_1 \groupcombine G_2$ is also a group (notice that privacy policies support group hierarchies),
\item a set of context variables $\mathcal{X}$, where each $X \in \mathcal{X}$ has a finite domain $D_X$ of possible values ranged over by $v_X$; $v$ ranges over the union $\bigcup_{X} D_X$.
\end{enumerate*}

The following can then be defined:
\begin{enumerate*}
\item a set of names $\mathcal{N}$ containing channel names and all the values of context variables,
\item a set of types $\mathcal{T}$ (ranged over by $T$), containing all basic types, all context variables, and every element of the form $\typeGroup{G}{T}$ (the intuition being that a name of type $\typeGroup{G}{T}$ can be used by processes that \enquote{belong} to group $G$ in order to exchange messages of type $T$).
\end{enumerate*}

Terms of the Privacy Calculus are defined in two levels, processes and systems:
{
\setlength{\columnsep}{26pt}
\begin{multicols}{2}
\begin{bnf}
P : Process ::=
| $\processnull$ : Empty
| $\processin{x}{x}{T} P$ : Input
| $\processout{x}{x} P$ : Output
| $\processsilent P$ : Silent
| $\processname{x}{T} P$ : Create channel
| $P \processpar P$ : Parallel
| $P \processchoice P$ : Choice
| $\processfullconditional{x}{v}{P}{P}$ : Conditional
| $\processrep P$ : Replication
;;
\end{bnf}
\begin{bnf}
S : System ::=
| $\systemnull$ : Empty
| $\systemname{x}{T} S$ : Create channel
| $S \systempar S$ : Parallel
| $\systemgroup{R} S$ : Bind group
| $\systempurpose{G}{u} P$ : Process lift
;;
\end{bnf}
\end{multicols}
}

Processes are standard in \piCalculus/: the empty process does nothing, the input $\processin{x}{y}{T} P$ receives $y$ of type $T$ via the channel $x$ (binding the name $y$ in $P$) and continues as $P$, the output $\processout{x}{y} P$ sends $y$ via the channel $x$ and continues as $P$, the silent process $\processsilent P$ does some unspecified internal work and then continues as $P$, the process $\processname{x}{T} P$ creates a channel $x$ of type $T$ (and binds the name $x$) in $P$, the parallel composition $P_1 \processpar P_2$ combines the two processes so that both run in parallel, the choice composition $P_1 \processchoice P_2$ combines the two processes so that only one will run, the conditional $\processfullconditional{x}{v}{P_1}{P_2}$ checks whether $x$ is equal to $v$ and if so continues as $P_1$, otherwise as $P_2$, and the replication $\processrep P$ behaves as $P \processpar \processrep P$. For brevity, define $\processeqconditional{x}{v}{P}$ as $\processfullconditional{x}{v}{P}{\processnull}$, $\processuneqconditional{x}{v}{P}$ as $\processfullconditional{x}{v}{\processnull}{P}$, $\prod_{i=1}^{n} P_i$ as $P_1 \processpar \ldots \processpar P_n$, and $\sum_{i=1}^{n} P_i$ as $P_1 \processchoice \ldots \processchoice P_n$. 

Systems annotate processes with high-level privacy information. The system $\systempurpose{G}{u}{P}$ declares that the process $P$ runs on behalf of group $G$ for the purpose $u$, the system $\systemgroup{R}{S}$ declares that the system $S$ runs for the role $R$ (in addition to any other groups declared in $S$), while the empty, name binding, and parallel systems are similar to the respective processes. For brevity, define $\prod_{i=1}^{n} S_i$ as $S_1 \systempar \ldots \systempar S_n$.

For the unambiguous treatment of bound names, CINNI~\cite{stehr2000} is employed: it adds indices to names, so that, for example, the term $\processname{x}{T} \processout{x}{y} \processname{x}{T} \processout{x}{y}\processnull$ is actually interpreted as $\processname{x_1}{T} \processout{x_1}{y} \processname{x_0}{T} \processout{x_0}{y}\processnull$. Thus, name substitution $\subst{b}{a}$ (substitute all free occurrences of $a$ with $b$) can be defined elegantly. For technical reasons, CINNI defines some operators that convert indices, such that $\shiftdown{a}{X}$, which decreases the indices of every $a$ in $X$ by 1 (not going below 0). In this paper, CINNI will be ignored, i.e. indices of channel names will always be omitted and the index 0 will be assumed for all channel names.

Two processes/systems that differ only in the selection of their bound names are called \alphaEquivalent/. Structural congruence (i.e. behavioural equivalence) of processes/systems, denoted $\equiv$, is defined as follows:
\begin{enumerate*}
\item \alphaEquivalent/ terms are congruent,
\item parallel/choice terms that differ only in the order of their operands are congruent (i.e. parallel and choice operators are associative and commutative),
\item repetitions of operands in choice is irrelevant (i.e. the choice operator is idempotent),
\item $P \processpar \processnull \equiv P$, $S \systempar \systemnull \equiv S$, $\processrep \processnull \equiv \processnull$, $\processname{x}{T}\processnull \equiv \processnull$ (both for processes and systems), $\systemgroup{G}{\processnull} \equiv \processnull$, $\systempurpose{G}{u}{\processnull} \equiv \systemnull$.
\end{enumerate*}

The operational semantics of the Privacy Calculus is defined in \cref{fig:pi-transitions}. It is a late labelled transition semantics comprising four kinds of labels: 
\begin{enumerate*}
\item silent, $\transitionsilent$,
\item input, $\transitionin{x}{y}$,
\item output, $\transitionout{x}{y}$,
\item bound output $\transitionboundout{x}{y}{T}$.
\end{enumerate*}
Notice that the \piCalculus/, and hence Privacy Calculus, is non-deterministic: multiple execution steps might be possible for a given term, in which case any one of them might be selected arbitrarily as the next to be executed. 

\begin{figure}[!hbt]
\begin{center}
\begin{minipage}{0.25\textwidth}\begin{equation}
\label{pi:In} 
\processin{x}{a}{T} P \transitionin{x}{a} P
\tag{In}
\end{equation}\end{minipage}
\begin{minipage}{0.25\textwidth}\begin{equation}
\label{pi:Out} 
\processout{x}{y}P \transitionout{x}{y} P
\tag{Out}
\end{equation}\end{minipage}
\begin{minipage}{0.2\textwidth}\begin{equation}
\label{pi:Silent} 
\processsilent P \transitionsilent P
\tag{Silent}
\end{equation}\end{minipage}
\begin{minipage}{0.3\textwidth}\begin{equation}
\label{pi:Repl} 
\AxiomC{$P \transition{l} P^\prime$}
\UnaryInfC{$\processrep P \transition{l} P^\prime \processpar \processrep P $}
\DisplayProof
\tag{Repl}
\end{equation}\end{minipage}
\begin{minipage}{0.35\textwidth}\begin{equation}
\label{pi:Choice} 
\AxiomC{$P_1 \transition{l} P_1^\prime$}
\UnaryInfC{$P_1 \processchoice P_2 \transition{l} P_1^\prime$}
\DisplayProof
\tag{Choice}
\end{equation}\end{minipage}
\begin{minipage}{0.4\textwidth}\begin{equation}
\label{pi:CondT} 
\AxiomC{$P \transition{l} P^{\prime}$}
\UnaryInfC{$\processfullconditional{x}{x}{P}{Q}  \transition{l}P^\prime $}
\DisplayProof
\tag{CondT}
\end{equation}\end{minipage}
\begin{minipage}{0.4\textwidth}\begin{equation}
\label{pi:CondF} 
\AxiomC{$Q \transition{l} Q^{\prime}$}
\AxiomC{$x \neq y$}
\BinaryInfC{$\processfullconditional{x}{y}{P}{Q}  \transition{l} Q^\prime $}
\DisplayProof
\tag{CondF}
\end{equation}\end{minipage}
\begin{minipage}{0.35\textwidth}\begin{equation}
\label{pi:ResGS} 
\AxiomC{$S \transition{l} S^\prime$}
\UnaryInfC{$\systemgroup{R}{S} \transition{l} \systemgroup{R}{S^\prime}$}
\DisplayProof
\tag{ResGS}
\end{equation}\end{minipage}
\begin{minipage}{0.45\textwidth}\begin{equation}
\label{pi:ResGP} 
\AxiomC{$P \transition{l} P^\prime$}
\UnaryInfC{$\systempurpose{G}{u}{P}  \transition{l} \systempurpose{G}{u}{P^\prime}$}
\DisplayProof
\tag{ResGP}
\end{equation}\end{minipage}
\begin{minipage}{0.4\textwidth}\begin{equation}
\label{pi:Open}
\AxiomC{$F \transitionout{x}{a_0} F^\prime$}
\UnaryInfC{$\processname{a}{T} F \transitionboundout{x}{a_0}{T} F^\prime$}
\DisplayProof
\tag{Open}
\end{equation}\end{minipage}
\begin{minipage}{0.5\textwidth}\begin{equation}
\label{pi:ResN}
\AxiomC{$F \transition{l} F^\prime$}
\AxiomC{$a_0 \notin \freenames({l})$}
\BinaryInfC{$\processname{a}{T} F \transition{\shiftdown{a}{l}}\processname{a}{T} F^\prime$}
\DisplayProof
\tag{ResN}
\end{equation}\end{minipage}
\begin{minipage}{0.4\textwidth}\begin{equation}
\label{pi:Com}
\AxiomC{$F_1 \transitionin{x}{a} F_1^\prime$}
\AxiomC{$F_2 \transitionout{x}{z} F_2^\prime$}
\BinaryInfC{$F_1 \processpar F_2 \transitionsilent (\subst{z}{a}{F_1^\prime}) \processpar F_2^\prime $}
\DisplayProof 
\tag{Comm}
\end{equation}\end{minipage}
\begin{minipage}{0.5\textwidth}\begin{equation}
\label{pi:Close}
\AxiomC{$F_1 \transitionin{x}{a} F_1^\prime$}
\AxiomC{$F_2 \transitionboundout{x}{b_n}{T} F_2^\prime$}
\BinaryInfC{$F_1 \processpar F_2 \transitionsilent \processname{b}{T}((\subst{b_n}{a}{F_1^\prime}) \processpar F_2^\prime) $}
\DisplayProof 
\tag{Close}
\end{equation}\end{minipage}
\begin{minipage}{0.4\textwidth}\begin{equation}
\label{pi:Par} 
\AxiomC{$F_1 \transition{l} F_1^\prime$}
\AxiomC{$\boundnames({l})\cap \freenames({F_2}) = \emptyset$}
\BinaryInfC{$F_1 \processpar F_2 \transition{l} F_1^\prime \processpar F_2$}
\DisplayProof
\tag{Par}
\end{equation}\end{minipage}
\begin{minipage}{0.35\textwidth}\begin{equation}
\label{pi:Congr} 
\AxiomC{$F_1\equiv F_2$}
\AxiomC{$F_2 \transition{l} F$}
\BinaryInfC{$F_1 \transition{l} F$}
\DisplayProof
\tag{Congr}
\end{equation}\end{minipage}
\end{center}
\caption{The rules of labelled transition semantics of the Privacy Calculus. $\freenames(X)$ is the set of free names of the term $X$, while $\boundnames(X)$ is the set of its bound names. Rules that contain the variable $F$ are applicable both to processes and systems.
}
\label{fig:pi-transitions}
\end{figure}

\section{Converting BPMN to Privacy Calculus}\label{sect:conversion}

\subsection{Main considerations regarding conversion}\label{sect:conversion-intro}

In the spirit of~\cite{puhlmann2005,puhlmann2006,boussetoua2015}, a BPMN Process will be converted to a Privacy Calculus term consisting of the concatenation of terms corresponding to every Flow Node (i.e. Event, Activity, or Gateway) within the Process. Flows will be converted to channels that serve for communication between processes: Sequence Flows will carry tokens, while Message Flows will carry messages (possibly containing data important to privacy policies).
Every Privacy Calculus process corresponding to a Flow Node will then have (roughly)\footnote{As~\cite[Section~4]{puhlmann2005} points out, \enquote{The description given applies only to basic control flow structures. Advanced structures require slightly different approaches.}; in fact, various of the patterns presented in~\cite{puhlmann2005,puhlmann2006} and in \cref{sect:conversion-elements} of this paper diverge (slightly or more radically) from the rough structure above. Moreover, \enquote{if a process representing \textins*{a Flow Node} can be triggered more than once, the replication operator must be used} and \enquote{a \textins{conditional} prefix \textins{after receiving the triggers} can be used to model global constraints like testing a cancellation flag} (the latter is in fact taken into account in the basic description given in~\cite{puhlmann2005}, but is left out here).} the following structure:
\begin{enumerate}
\item begin with receiving tokens via its incoming Sequence Flows,
\item continue with receiving Messages via its incoming Message Flows,
\item do any work specific to its type,
\item send messages via its outgoing Message Flows,
\item pass token(s) to its outgoing Sequence Flow(s); some outgoing Sequence Flows might be affected by conditions, hence only be triggered conditionally.
\end{enumerate}
This design choice is influenced by~\cite{puhlmann2005}:
\blockcquote[Section~4]{puhlmann2005}{A generic process can have $m$ incoming triggers \textelp{} and $o$ outgoing triggers. \textelp{} After the input prefixes have been triggered \textelp{} First, the functional perspective of the activity is represented as an unobservable action. Second, the process can trigger other processes by output prefixes.}. Here, we have \enquote{unfolded} a small part of the unobservable action so as to accommodate Message Flows and considered that some outgoing Sequence Flows are only conditionally triggered.

In order to accommodate Sequence Flows, the set $\mathcal{T}$ of types is presumed to contain a special type $\tokenType$ of tokens. For simplicity, values of this type will always be denoted by $\tokenValue$, assuming that this name is not used for any other element.

Notice that
\blockcquote[Section~7.2]{bpmn}{a token does not traverse a Message Flow since it is a Message that is passed down a Message Flow (as the name implies)}. Hence, for every Message Flow, the name and type (as an element of $\mathcal{T}$) of the Message needs to be known; types of Messages are not a part of BPMN and, as stressed in \cref{sect:bpmn}, names of Messages will need to be provided externally to BPMN.

\subsection{Flow patterns}\label{sect:conversion-patterns}

In \cite{puhlmann2005,puhlmann2006}, various patterns common to business processes are identified and their conversion to \piCalculus/ processes is discussed. We will review a few here (Sequence, Parallel split, Exclusive choice, Synchronisation, N–out–of–M–join) and present some variations of them (Choice, $n$–out–of–$n$ synchronisation, $m$–out–of–$n$ synchronisation).

While discussing flow patterns, the following notation will be adopted:
\begin{enumerate*}
\item Instead of BPMN Flow Nodes and Flows, arbitrary nodes and edges (in the graph theoretic sense) will be considered.
\item The conversion of a node $X$ into a Privacy Calculus term will be denoted $\convertToPrivacyCalculus{X}$.
\item Since only the initial and/or final behaviour of $X$ will be of interest, there will be a part of $\convertToPrivacyCalculus{X}$ that will be irrelevant to this discussion (in fact, it will depend on what kind of BPMN element $X$ is); this will be denoted by $X^\prime$.
\end{enumerate*} 

\paragraph{Sequence:}
The simplest pattern, defined in~\cite[Section~4.1]{puhlmann2005}. 

Suppose that node $A$ has a unique outgoing edge $f$.
Then, $A$, when it has finished its work, needs only trigger the next node by sending a token via $f$, i.e.
$
\convertToPrivacyCalculus{A} \defEq A^\prime. \processout{f}{\tokenValue}\processnull
$.

Similarly, suppose that node $B$ has a unique incoming edge $f$.
Then, $B$ waits until it receives the token and then starts its own work, i.e.
$
\convertToPrivacyCalculus{B} \defEq \processin{f}{\tokenValue}{\tokenType} B^\prime
$.

\subsubsection{Outgoing}

Suppose that node $A$ has multiple outgoing edges $f_1, \ldots, f_n$ ($n \geq 2$) to other nodes.

\paragraph{Parallel split:} 
In this pattern, defined in~\cite[Section~4.1]{puhlmann2005}, $A$ triggers all of its outgoing edges in parallel. For $n = 2$, this can be achieved with
$
\convertToPrivacyCalculus{A} \defEq A^\prime. (\processout{f_1}{\tokenValue}\processnull  \processpar \processout{f_2}{\tokenValue}\processnull)
$.
This can be generalised to
$
\convertToPrivacyCalculus{A} \defEq A^\prime. \prod_{i=1}^{n}\processout{f_i}{\tokenValue}\processnull
$ .

\paragraph{Exclusive choice:}
In this pattern, defined in~\cite[Section~4.1]{puhlmann2005}, $A$ triggers exactly one of its outgoing edges. For $n = 2$, this can be achieved with
$
\convertToPrivacyCalculus{A} \defEq A^\prime. (\processout{f_1}{\tokenValue}\processnull  \processchoice \processout{f_2}{\tokenValue}\processnull)
$.
This can be generalised to
$
\convertToPrivacyCalculus{A} \defEq A^\prime. \sum_{i=1}^{n}\processout{f_i}{\tokenValue}\processnull
$ .

\subsubsection{Incoming}

Suppose that node $B$ has multiple incoming edges $f_1, \ldots, f_n$ ($n \geq 2$) from other nodes.

\paragraph{Choice:}
In this pattern, $B$ waits for any of its incoming edges to be triggered and then starts. Input from the rest of the edges is disregarded. For $n = 2$, this can be achieved with $\convertToPrivacyCalculus{B} \defEq \processin{f_1}{\tokenValue}{\tokenType}B^\prime \processchoice \processin{f_2}{\tokenValue}{\tokenType}B^\prime$. This can be generalised to 
$
\sum_{i=1}^{n} \processin{f_i}{\tokenValue}{\tokenType}B^\prime
$.

If multiple incoming edges can be activated, then the choice pattern will process only one of them, leaving the rest \enquote{hanging}. Depending on the situation at hand, this might be alleviated (if needed) either by creating a new copy of $B$ for each incoming trigger (similarly to the \emph{Multi-merge} pattern of~\cite[Section~4.2]{puhlmann2005}) or by the \emph{1-out-of-$n$ synchronisation} pattern below.

\paragraph{Synchronisation:}
In this pattern, defined in~\cite[Section~4.1]{puhlmann2005}, $B$ waits for all of its incoming edges to be triggered\textemdash in a predefined order, however\textemdash before it starts. For $n = 2$, this can be achieved with $\convertToPrivacyCalculus{B} \defEq \processin{f_1}{\tokenValue}{\tokenType}\processin{f_2}{\tokenValue}{\tokenType}B^\prime$. This can be generalised to 
$
\processin{f_1}{\tokenValue}{\tokenType}\ldots\processin{f_n}{\tokenValue}{\tokenType}B^\prime
$.

We will not use this pattern in \cref{sect:conversion-elements}, opting for the $n$-out-of-$n$ and $m$-out-of-$n$ variants below. The reason is twofold. First, notice that, in general, this pattern might create deadlock issues; for instance, consider $A$ triggering both $C$ and $B$ (via $f_2$) and $C$ triggering $B$ (via $f_1$): if the outgoing pattern used by $A$ waits for $f_2$ to be consumed before triggering $C$, then $B$ will never be executed. Moreover, even if such deadlocks are guaranteed to be impossible, $\convertToPrivacyCalculus{B}$ will be behaviourally different depending on the order the $f_i$ are written; this asymmetry might be undesired in applications such as the one of \cref{sect:converter-app} (e.g. it might complicate unit testing, since a single input will have multiple non-equivalent correct outputs).

\paragraph{$n$–out–of–$n$ synchronisation:}
In this pattern (similar to \emph{N-out-of-M-join} of~\cite[Section~4.2]{puhlmann2005}), $B$ (run on behalf of group $G$) waits for all of its incoming edges to be triggered before it starts, consuming every trigger as it arrives. For $n = 2$, this can be achieved with
\[
\convertToPrivacyCalculus{B} \defEq \processname{h}{G[\tokenType]}
\processin{h}{\tokenValue}{\tokenType}
\processin{h}{\tokenValue}{\tokenType} 
B^\prime\processpar
\processin{f_1}{\tokenValue}{\tokenType}\processout{h}{\tokenValue}\processnull\processpar
\processin{f_2}{\tokenValue}{\tokenType}\processout{h}{\tokenValue}\processnull,
\]
where $h$ must not be free in $B^\prime$. This can be generalised to
\begin{equation*}
\convertToPrivacyCalculus{B} \defEq
\processname{h}{G[\tokenType]}
\underbrace{
\processin{h}{\tokenValue}{\tokenType}
\ldots
\processin{h}{\tokenValue}{\tokenType}
}_\text{$n$ times}
B^\prime
\processpar 
\prod_{i=1}^n
\processin{f_i}{\tokenValue}{\tokenType}\processout{h}{\tokenValue}\processnull\ .
\end{equation*}
The drawback of this pattern is that it creates a fresh name. Applications such as the one of \cref{sect:converter-app} need to select a name not among the free names of $B^\prime$. Moreover, it might complicate unit testing: either the selected name must be known when writing tests or \alphaEquivalence/ must be tested instead of equality.

\paragraph{$m$–out–of–$n$ synchronisation:}

In this pattern, generalising the previous one, $B$ (run on behalf of group $G$) waits for exactly $m \leq n$ of its incoming edges to be triggered before it starts, consuming however all $n$ triggers as they arrive. For $n = 2$ and $m = 1$, this can be achieved with
\[
\begin{aligned}
\convertToPrivacyCalculus{B} \defEq 
&\processname{h}{G[\tokenType]}\processname{r}{G[\tokenType]}\\
&
\processin{r}{\tokenValue}{\tokenType}B^\prime
\processpar
\processin{f_1}{\tokenValue}{\tokenType}\processout{h}{\tokenValue}\processnull\processpar
\processin{f_2}{\tokenValue}{\tokenType}\processout{h}{\tokenValue}\processnull\processpar
\processin{h}{\tokenValue}{\tokenType}
\processout{r}{\tokenValue}
\processin{h}{\tokenValue}{\tokenType}
\processnull,
\end{aligned}
\]
where $h$ and $r$ must not be free in $B^\prime$. This can be generalised to
\[
\begin{aligned}
\convertToPrivacyCalculus{B} \defEq 
&\processname{h}{G[\tokenType]}\processname{r}{G[\tokenType]} \processin{r}{\tokenValue}{\tokenType}B^\prime
\processpar
\prod_{i=1}^{n}\processin{f_i}{\tokenValue}{\tokenType}\processout{h}{\tokenValue}\processnull \processpar \\
&
\underbrace{\processin{h}{\tokenValue}{\tokenType}\ldots\processin{h}{\tokenValue}{\tokenType}}_\text{$m$ times}
\processout{r}{\tokenValue}
\underbrace{\processin{h}{\tokenValue}{\tokenType}\ldots\processin{h}{\tokenValue}{\tokenType}}_\text{$n - m$ times}
\processnull\ .
\end{aligned}
\]

The drawback of this pattern, as with the previous one, is that it creates fresh names.
This pattern might be useful for the conversion of some kinds of Complex Gateways~\cite[Section~10.6.5]{bpmn} (e.g. those operating on the rule that \enquote{three out of five incoming Sequence Flows are needed to activate the Gateway}), which however are not considered here.
\cite[Section~4.2]{puhlmann2005} defines the similar pattern \emph{N–out–of–M–join}, which recursively restarts $\convertToPrivacyCalculus{B}$ after having consumed all of the $n$ input triggers.

\subsection{Conversion of diagram elements}\label{sect:conversion-elements}

This section is the gist of this paper. In \cref{sect:conversion-processes,sect:conversion-events-start,sect:conversion-events-end,sect:conversion-events-intermediate,sect:conversion-gateways-parallel,sect:conversion-gateways-exclusive,sect:conversion-tasks,sect:conversion-subprocesses,sect:conversion-participants,sect:conversion-collaborations}, the conversion of every supported kind of BPMN element to Privacy Calculus is discussed.
For any BPMN element $N$, the corresponding Privacy Calculus term will be denoted $\convertToPrivacyCalculus{N}$.

\subsubsection{Start Events}\label{sect:conversion-events-start}

The most generic form of a Start Event $N$ is for it to have 
\begin{enumerate*}
\item no incoming Sequence Flows, 
\item multiple outgoing Sequence Flows $f_1, \ldots, f_k$, $1 \leq k$, and
\item (if it is a Message Start Event) multiple incoming Message Flows $E_1, \ldots, E_l$, $1 \leq l$, each carrying a message $m_i$ of type $T_i$.
\end{enumerate*}

According to~\cite[Section~10.5.2]{bpmn}, each Message Flow targeting a Start Event represents an instantiation mechanism (a trigger) for the Process; only one of the triggers is required to start a new Process. Thus, $\convertToPrivacyCalculus{N}$ shall start with a \emph{Choice} pattern among the Message Flows. Also, according to~\cite[Section~10.5.2]{bpmn}, if multiple Sequence Flows originate at a Start Event, then they are considered as parallel paths; thus the \emph{Parallel split} pattern shall be used. Hence, $\convertToPrivacyCalculus{N}$ will be 
\[
\sum_{i=1}^l \left(\processin{E_i}{m_i}{T_i} \processname{\tokenValue}{\tokenType} \prod_{j=1}^k \processout{f_j}{t} \processnull \right),
\]
that is, $N$ waits for any $E_i$ to pass a message and then, being a Start Event, generates a token. It has no further internal work to do, so it triggers all of its outgoing Sequence Flows in parallel using the \emph{Parallel split} pattern.
Of course, in case there are no incoming Message Flows (i.e. the Start Event is not a Message Event), $\convertToPrivacyCalculus{N}$ can be simplified to $\processname{\tokenValue}{\tokenType}\prod_{j=1}^k \processout{f_j}{\tokenValue} \processnull$.

In case that a Message Start Event is part of a single Process (i.e. not in a Collaboration) or the modeller has failed to provide Message Flows, one \enquote{phantom} Message Flow can be assumed and the conversion can then still proceed as above.

\subsubsection{End Events}\label{sect:conversion-events-end}

The most generic form of an End Event $N$ is for it to have 
\begin{enumerate*}
\item multiple incoming Sequence Flows $e_1, \ldots, e_k$, $1 \leq k$, 
\item no outgoing Sequence Flows, and
\item (if it is a Message End Event) multiple outgoing Message Flows $F_1, \ldots, F_l$, $1 \leq l$, each carrying a message $m_i$ of type $T_i$; it is assumed here that the messages are generated within the Event.
\end{enumerate*}
Suppose that the Process containing $N$ runs for group $G$.

Contrary to~\cite[Section~10.5.3]{bpmn}, if multiple Sequence Flows converge into an End Event, they will be required to be parts of parallel paths; then, according to~\cite[Section~10.5.3]{bpmn}, \enquote{the tokens will be consumed as they arrive}. Hence, the End Event starts with a \emph{$k$-out-of-$k$ synchronisation} pattern. Afterwards, \blockcquote[Section~10.5.3]{bpmn}{Each Message Flow leaving the End Event will have a Message sent when the Event is triggered.}, which indicates a \emph{Parallel split} of Message Flows. Hence, for $k > 1$, $\convertToPrivacyCalculus{N}$ will be
\[
\processname{h}{\typeGroup{G}{\tokenType}} \underbrace{\processin{h}{\tokenValue}{\tokenType} \ldots \processin{h}{\tokenValue}{\tokenType}}_\text{$k$ times} D \processpar \prod_{i=1}^{k} \processin{e_i}{\tokenValue}{\tokenType}\processout{h}{\tokenValue}\processnull,
\]
where $D$ is
$
\prod_{j=1}^{l}\processname{m_j}{T_j} \processout{F_j}{m_j}\processnull
$
for Message End Events and $\processnull$ otherwise. For $k = 1$, $\convertToPrivacyCalculus{N}$ can be simplified to $\processin{e_1}{\tokenValue}{\tokenType} D$ (a \emph{Sequence} pattern). If the End Event is part of a Sub-Process, the $\processnull$ at the end of $D$ is replaced by a \emph{Parallel split} pattern of the Sequence Flow(s) outgoing from the Sub-Process.

In case that a Message End Event is part of a single Process (i.e. not in a Collaboration) or the modeller has failed to provide Message Flows, one \enquote{phantom} Message Flow can be assumed and the conversion can then still proceed as above.

\subsubsection{Intermediate Events}\label{sect:conversion-events-intermediate}

Recall that only Message Intermediate Events are considered in this paper.

Every Message Intermediate Event can be the source or target (depending on whether the Event is catching or throwing) of at most one Message Flow~\cite[Section~10.5.4]{bpmn}. Moreover, contrary to~\cite[Section~10.5.4]{bpmn}, here it will be assumed that every Intermediate Event has at most (hence, exactly) one incoming Sequence Flow. The most generic form of a Message Intermediate Event $N$ is hence for it to have 
\begin{enumerate*}
\item one incoming Sequence Flow $e_1$, 
\item multiple outgoing Sequence Flows $f_1, \ldots, f_n$, $1 \leq n$, and
\item (if it is a Message Intermediate Catch Event) one incoming Message Flow $E$, carrying a message $m$ of type $T$,
\item (if it is a Message Intermediate Throw Event) one outgoing Message Flow $F$, carrying a message $m$ of type $T$; it is assumed here that the outgoing message is generated within the Event.
\end{enumerate*}

According to~\cite[Section~10.5.4]{bpmn}, if multiple Sequence Flows originate at an Intermediate Event, then they are considered as parallel paths. Hence the event can use the \emph{Sequence} (for incoming) and \emph{Parallel split} (for outgoing) patterns and $\convertToPrivacyCalculus{N}$ is
\begin{align*}
& \processin{e_1}{\tokenValue}{\tokenType} \processin{E}{m}{T}\prod_{j=1}^{n} \processout{f_i}{\tokenValue} \processnull && \text{for Catch Events,}\\
&\processin{e_1}{\tokenValue}{\tokenType} \processname{m}{T}\processout{F}{m}\prod_{j=1}^{n} \processout{f_i}{\tokenValue} \processnull &&\text{for Throw Events.}
\end{align*}

Notice that this is a simplified conversion. In fact, \blockcquote[Section~10.5.4]{bpmn}{if another token arrives from the same path or another path, then a separate instance of the Event will be created}. However, multiple instances of Events will not be tackled here, since that would be quite more complicated (as \cite[Section~4.2]{puhlmann2005} points out, \enquote{by using the replication operator to create multiple copies of a process $D$, all processes that are triggered by $D$ must also support replication and so on. This also refers to all other patterns that create multiple copies by replication.}) and of minimal interest regarding privacy protection.

In case that a Message Intermediate Event is part of a single Process (i.e. not in a Collaboration) or the modeller has failed to provide a Message Flow for the Event, one \enquote{phantom} Message Flow can be assumed and the conversion can then still proceed as above.

\subsubsection{Parallel Gateways}\label{sect:conversion-gateways-parallel}

Every Parallel Gateway $N$ has
\begin{enumerate*}
\item $1 \leq k$ incoming Sequence Flows $e_1, \ldots, e_k$,  and
\item $1 \leq n$ outgoing Sequence Flows $f_1, \ldots, f_n$.
\end{enumerate*}
Since Gateways have no internal operation, Parallel Gateways can be modelled using only the \emph{$k$-out-of-$k$ synchronisation} and \emph{Parallel split} patterns, i.e. for $k > 1$, $\convertToPrivacyCalculus{N}$ will be
\[
\processname{h}{\typeGroup{G}{\tokenType}} \underbrace{\processin{h}{\tokenValue}{\tokenType} \ldots \processin{h}{\tokenValue}{\tokenType}}_\text{$k$ times} \left(\prod_{i=1}^{n} \processout{f_i}{\tokenValue}\processnull\right) \processpar \prod_{i=1}^{k} \processin{e_i}{\tokenValue}{\tokenType}\processout{h}{\tokenValue}\processnull,
\]
and for $k = 1$ it will be simplified to 
$
\processin{e_1}{\tokenValue}{\tokenType}  \left(\prod_{i=1}^{n} \processout{f_i}{\tokenValue}\processnull\right)
$.

\subsubsection{Exclusive Gateways}\label{sect:conversion-gateways-exclusive}

Under the simplifying conventions introduced in \cref{sect:bpmn-gateways}, an Exclusive Gateway $N$ can only have the following form:
\begin{enumerate*}
	\item multiple incoming Sequence Flows $e_1, \ldots, e_k$, $2 \leq k$, at most one of which will be triggered, and
	\item one outgoing Sequence Flow $f_1$.
\end{enumerate*}
Hence, a \emph{Choice} pattern for input and a \emph{Sequence} pattern for output shall be adequate and $\convertToPrivacyCalculus{N}$ can be
$
\sum_{i=1}^{k}\processin{e_i}{\tokenValue}{\tokenType}\processout{f_1}{\tokenValue}\processnull
$ .

\subsubsection{Tasks}\label{sect:conversion-tasks}
Under the simplifying conventions introduced in \cref{sect:bpmn-activities,sect:bpmn-flows}, the most generic form of a Task $N$ is for it to have 
\begin{enumerate*}
\item one incoming Sequence Flow $e_1$, 
\item 0 or 1 incoming Message Flows $E$ (1 in case the Task is a Receive Task), carrying a message $m_E$ of type $T_E$, 
\item multiple outgoing Sequence Flows $f_1, \ldots, f_n$, $1 \leq n$, where, if the Task is a Receive Task, each $f_i$ might have a condition $c_i$ attached ($c_i$ compares $m_E$ to some constant value $v_i$ via $o_i$, where $o_i$ can be either $=$ or $\neq$),
\item 0 or 1 outgoing Message Flows $F$ (1 in case the Task is a Send Task), carrying a message $m_F$ of type $T_F$; it is assumed here that the outgoing message is generated within the Task.
\end{enumerate*}

Outgoing Sequence Flows of Tasks need a \emph{Parallel split} pattern, since \blockcquote[Section~10.3]{bpmn}{if there are multiple outgoing Sequence Flows, then this means that a separate parallel path is being created for each Sequence Flow}.
For each other Flow kind, since at most one item exists, the \emph{Sequence} pattern suffices.

In~\cite[Section~4]{puhlmann2005}, it is argued that \enquote{a process that represents an activity must have a functional part represented by $\tau$}; recall that the $\processsilent$ prefix in Privacy Calculus encodes that some unspecified internal work is performed. This is indeed compatible with the fact that Tasks are used in BPMN \blockcquote[Section~7.3.2]{bpmn}{when the work \textelp{} is not broken down to a finer level of \textelp{} detail}. 

Hence, $\convertToPrivacyCalculus{N}$ will be
\[
\processin{e_1}{\tokenValue}{\tokenType} \processin{E}{m_E}{T_E} \processsilent \processname{m_F}{T_F}\processout{F}{m_F} \prod_{i=1}^{n}\processconditional{m_E}{o_i}{v_i}{\processout{f_i}{\tokenValue} \processnull}
\]
where $\processin{E}{m}{T}$ shall be omitted if there is no $E$, $\processname{m}{T}\processout{F}{m}$ shall be omitted if there is no $F$, and $\processconditional{m_E}{o_i}{v_i}{}$ shall be omitted if there is no $c_i$.
For a Send/Receive Task, the $\processsilent$ shall be omitted, since there is no internal work other than sending/receiving the Message (\cite[Section~10.3.3]{bpmn} stresses that once the Message has been sent/received, the Task is completed). 

Notice that this is a simplified conversion. In fact, similarly to Intermediate Events, \blockcquote[Section~10.3]{bpmn}{if another token arrives from the same path or another path, then a separate instance of the Activity will be created}. This will not be considered here, with a same rationale as for Intermediate Events.

In case that a Receive/Send Task is part of a single Process (i.e. not in a Collaboration) or the modeller has failed to provide a Message Flow for the Task, one \enquote{phantom} Message Flow (incoming for Receive, outgoing for Send) can be assumed and the conversion can then still proceed as above.

\subsubsection{Processes}\label{sect:conversion-processes}

As already mentioned in \cref{sect:conversion-intro}, a BPMN Process will be converted to a Privacy Calculus term consisting of the concatenation of Privacy Calculus subprocesses corresponding to every Flow Node within the Process. In fact, since we are interested in privacy protection, a top-level Process (i.e. not part of a Collaboration and not a Sub-Process) must be decorated with a group $G$ that runs the Process and a purpose $u$ for which it is run, so that it can be checked for compliance to privacy policies.

Notice that \blockcquote[Section~10.5.2]{bpmn}{each Start Event is an independent Event}, hence in case of multiple Start Events in the same Process (something permitted but not recommended in BPMN~\cite[Section~10.5.2]{bpmn}), the first one to be triggered invalidates (for the Process instance that is created) the rest.

Consider a Process $N$. Let $E_N$ be the set of Start Events of $N$. For every Start Event $E \in E_N$, let $A_E$ be the set of Flow Nodes (including itself) that are accessible (in the graph-theoretic sense) via Sequence Flows from $E$. Let also $S$ be the set of $m$ Sequence Flows which connect the nodes of $\bigcup_{E \in E_N} A_E$; $N$ must bind the names of the channels corresponding to the Flows in order to prevent usage from the outside.

Given the considerations above, $\convertToPrivacyCalculus{N}$ will be the Privacy Calculus system
\[
\systempurpose{G}{u}{\underbrace{\processname{f_1}{\typeGroup{G}{\tokenType}}\ldots\processname{f_m}{\typeGroup{G}{\tokenType}}}_{\text{for all $f_i \in S, i = 1,\ldots, m$}}\sum_{E \in E_N} \prod_{A \in A_E} \convertToPrivacyCalculus{A}}.
\]

\subsubsection{Sub-Processes}\label{sect:conversion-subprocesses}

Under the conventions of \cref{sect:bpmn-activities,sect:bpmn-flows}, the most generic form of a Sub-Process $N$ is for it to have 
\begin{enumerate*}
\item one incoming Sequence Flow $e_1$, 
\item multiple outgoing Sequence Flows $f_1, \ldots, f_n$, $1 \leq n$, 
\item a non-empty set $M$ of Flow Nodes and Flows within it.
\end{enumerate*}

Let $F$ be the process $\prod_{i=1}^{n}\processout{f_i}{\tokenValue}\processnull$ (\emph{Parallel split} of the outgoing Sequence Flows). Then $\convertToPrivacyCalculus{N}$ will be 
$
\processin{e_1}{\tokenValue}{\tokenType}\convertToPrivacyCalculus{M}
$,
where $M$ is converted as a Process (\cref{sect:conversion-processes}), with the exceptions that
\begin{enumerate*}
\item as noted in \cref{sect:conversion-events-end}, End Events of Sub-Processes are converted in a special manner: as their final step, instead of a plain $\processnull$, they contain $F$, thus shifting flow control back to the process that contains $N$,
\item a Sub-Process is not decorated with group/purpose information: it is considered to run for the same group and purpose as the Process containing it.
\end{enumerate*}
If $N$ is a multi-instance (parallel) Sub-Process, then $\convertToPrivacyCalculus{N}$ is
$
\processin{e_1}{\tokenValue}{\tokenType}\processrep\convertToPrivacyCalculus{M}
$.

\subsubsection{Participants}\label{sect:conversion-participants}

A Participant $N$ has, in general, one of the two following structures:
\begin{itemize}
\item It is either a Process $M$ with the (optional) information of a group/user $G$ that runs it. If we make this information required and also require a purpose $u$, then $\convertToPrivacyCalculus{N}$ can be the Privacy Calculus system $\systempurpose{G}{u}{\convertToPrivacyCalculus{M}}$,
\item Or it is just a group/user $G$ (again, the name is optional) depicted as a \enquote{black box}. In this case, we can use a variable $P_G$ for the Privacy Calculus process, require a group and a purpose, and set $\convertToPrivacyCalculus{N}$ to be the Privacy Calculus system $\systempurpose{G}{u}{P_G}$.
\end{itemize}

\subsubsection{Collaborations}\label{sect:conversion-collaborations}
A Collaboration $N$ is a non-empty collection of Participants $M_1, \ldots, M_n$, $1 \leq n$, of groups $G_1, \ldots G_n$, along with some Message Flows $F_1, \ldots, F_k$, $0 \leq k$, each $F_i$ carrying a Message of type $T_i$ between two Participants pertaining to groups $G_{i,1}$ and $G_{i,2}$. In addition to containing Participants, the converted term of the Collaboration needs to bind the Message Flows (for exactly the same reasons as Process binds Sequence Flows) and declare the group combinations in use. Thus, $\convertToPrivacyCalculus{N}$ is
\[
\systemgroup{G_{1,1}\groupcombine G_{1,2}}{
\ldots
\systemgroup{G_{k,1}\groupcombine G_{k,2}}{
\processname{F_1}{\typeGroup{G_{1,1}\groupcombine G_{1,2}}{T_1}} \ldots \processname{F_k}{\typeGroup{G_{k,1}\groupcombine G_{k,2}}{T_k}} \prod_{i=1}^{n} \convertToPrivacyCalculus{M_i}
}}.
\]

\subsection{A tool that automates the conversion}\label{sect:converter-app}

\begin{figure}[!hbt]
\begin{center}
\includegraphics[width=0.95\textwidth]{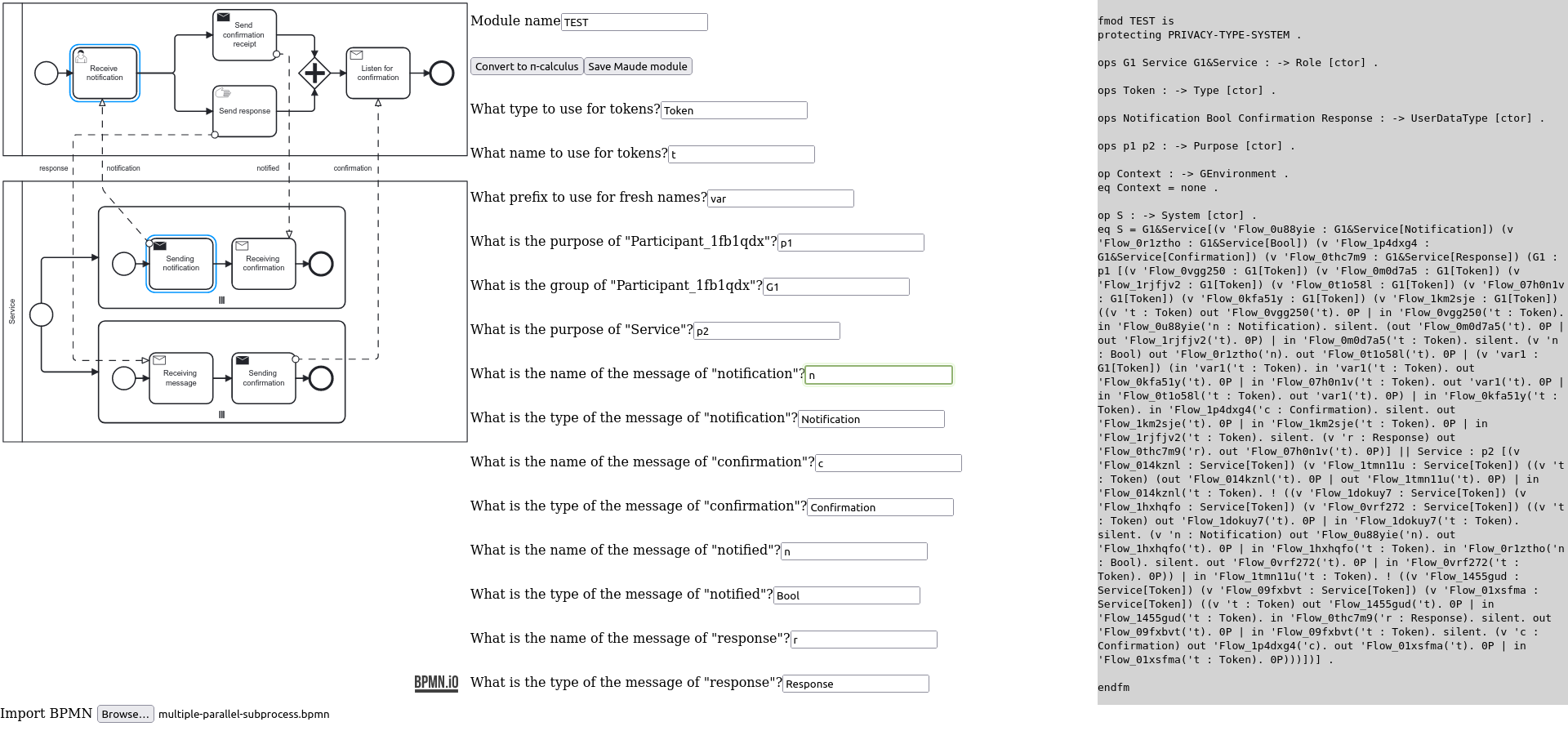}
\caption{A screenshot of the web app presented in \cref{sect:converter-app}. At the left, the imported diagram is shown and a new one can be uploaded. The middle part contains the extra information that the user needs to fill; for ease of use, the relevant Flow Nodes are highlighted when the user selects a question. The right part contains the Maude module created by the app (or the latest error that occurred).}
\label{fig:converter-app}
\end{center}
\end{figure}

An open source tool that automates the conversion has been implemented and its source code is available at~\cite{bpmn-to-privacy-converter-code}.
It is a simple web application written in HTML and Javascript (ES2022 dialect), which can run in modern web browsers. A screenshot of the app in use is shown in \cref{fig:converter-app}.

First, a BPMN XML diagram is imported. It must adhere to the assumptions mentioned in this paper.
The tool uses the open source library \href{https://github.com/bpmn-io/bpmn-js}{\texttt{bpmn-js}} to parse BPMN XML diagrams\footnote{The authors of \texttt{bpmn-js} have also created the app \href{https://bpmn.io}{bpmn.io} that can be used for creating and editing BPMN XML diagrams.}.

Afterwards, the app asks for any extra info needed.
It always requests the names to be used as 
\begin{enumerate*}
\item the type of tokens, 
\item the value of tokens,
\item prefix of fresh names;
\end{enumerate*}
this avoids using predefined values. Moreover, it asks for purposes and missing groups of Processes/Participants, and details (name, type, and, if \enquote{phantom}, channel) of Messages carried by (actual or \enquote{phantom}) Message Flows.

Finally, a Maude module compatible with~\cite{pitsiladis2018} is created. It contains a Privacy Calculus system \texttt{S} built using the conversions defined in \cref{sect:conversion}. In order for Maude to parse it correctly, all groups, types, purposes, values of user data types, and process/system variables used in \texttt{S} are first defined as terms of the module. For technical reasons having to do with the type checking algorithm of Privacy Calculus, the module also contains the context of \texttt{S}, i.e. a formal term specifying the types of free names of \texttt{S}. The Maude module can then be saved to a file and imported in the tool of~\cite{pitsiladis2018} for further processing.

\section{Conclusion and future work}\label{sect:conclusion}

We have presented how some basic elements of BPMN diagrams can be converted into Privacy Calculus and have provided a tool which can perform the conversion automatically and export it in a form compatible with the Maude formalisation of Privacy Calculus in~\cite{pitsiladis2018}.

As detailed in the previous sections, a significant subset of BPMN was not considered in this paper (and in the tool), but is required for detailed modelling of business processes. It is a matter of future work to integrate more aspects of BPMN.
For instance, supporting BPMN Data Objects and Data Stores would significantly boost the level of expressiveness; this can be achieved by taking advantage of some versions of Privacy Calculus that contain operators for data storage and retrieval~\cite{kouzapas2017,vanezi2020}.

A more accurate conversion of some Events and Activities could be achieved by carefully converting Flow Nodes to account for re-triggering. This might not be quite important \emph{as far as privacy is concerned}\footnote{At least at the current level of maturity of the Privacy Calculus ecosystem: if policies became more expressive (e.g. if they placed restrictions on \emph{how much} or \emph{how often} data is processed), then consideration of re-triggering would be required.}, since a business process is compliant iff every path is compliant, hence re-triggering a path or triggering another one is irrelevant. Of course, sound conversion\footnote{The BPMN standard does not define a formal semantics for diagrams, resting only on the treatment of tokens. Multiple researchers have proposed formalisations of (parts of) BPMN semantics in different frameworks: a few are~\cite{dijkman2008,dijkman2010,borger2011,wong2011,lam2012,elhichami2015,corradini2016,corradini2022,debrock2024}. Hence, soundness is in general not uniquely determined, since it depends on the selected formalisation.} is important for other reasons.

The toolchain of Privacy Calculus can (and should) also be complemented with more tools. For instance, a workflow for proving compliance of programs to policies would require tools that aid in the declaration of privacy policies, either by offering GUIs for the policy language of Privacy Calculus or by converting from more user-friendly frameworks.

\bibliographystyle{eptcs}
\bibliography{from2024-paper}
\end{document}